\documentclass[a4paper]{jpconf}
\usepackage{graphicx}
\usepackage[utf8]{inputenc}
\usepackage{amsmath}
\usepackage{amsfonts}
\usepackage{amssymb}
\usepackage{graphicx}
\usepackage{enumerate}
\usepackage[english]{babel}
\begin{document}
\title{Effects of the direct light in the surface detectors (SD) of
the Pierre Auger Observatory and their change in time}

\author{\small Pedro Alfonso Valencia Esquipula$^a$, Karen Salom\'e Caballero Mora$^a$ $^b$.}

\address{\small $^a$Facultad de Ciencias en F\'isica y Matem\'aticas (FCFM). Universidad Autónoma de Chiapas (UNACH), Chiapas, Mexico.$^b$Mesoamerican Centre for Theoretical Physics (MCTP).}

\ead{pa.valenciaesquipula@gmail.com}

\begin{abstract}
Cosmic Rays (CR) are particles which come to the earth from Universe. Their origin and production mechanisms are still unknown. The Pierre Auger Observatory is located in Mendoza, Argentina. It is dedicated to the study of CR. When CR arrive to the earth's atmosphere they produce a shower of secondary particles called \textit{air shower}.  The surface detector (SD) of the Pierre Auger Observatory consists of tanks full of pure water, where CR produce \textit{Cherenkov radiation}, when going through them. This light is detected by three photomultiplier tubes (PMT) located on the top of each tank. Depending of the angle of arrival direction of the primary CR, each PMT is able to register different signal than the other. The goal of this study is to look at these effects of direct light on the PMT's to explore if they change in time. The obtained results may give information about the physical status of the tanks in order to monitor the work of the SD, and to estimate possible systematic effects on the measurements. The current results of this study are shown.  
\end{abstract}

\section{Introduction}
The Pierre Auger Observatory measures Cosmic Rays within an area of 3000 $km^2$. The design calls for 1660 water-Cherenkov detectors, arranged on a triangular grid, with the sides of the triangles measuring 1.5 $km$, overlooked by four optical stations, each containing six telescopes, designed to detect air-fluorescence light (Fig. 1). The water-tanks(stations) respond to the particle component (mainly muons, electrons and positrons)
and the fluorescence cameras measure the emission from atmospheric nitrogen that is excited by the charged particles of the shower as they traverse the atmosphere. Both techniques, already used for many years to study extensive air showers, are brought together in a ‘hybrid’ detector to observe showers simultaneously with different techniques[1]. The array of water-tanks is known as the surface detector (SD) while the optical stations form the fluorescence detector, are known as fluorescence detector (FD).
\begin{center}
\includegraphics[width=.7\textwidth]{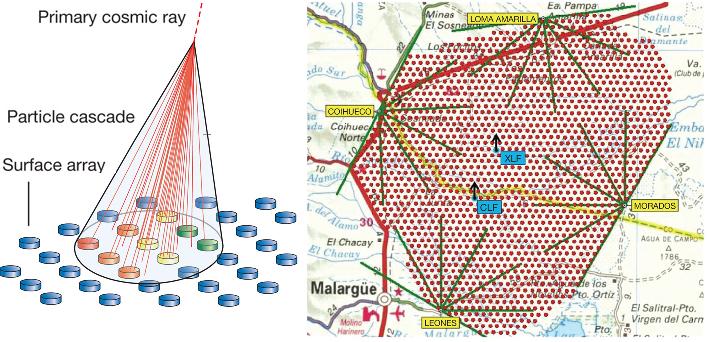}\\
{\small\textbf{Figure 1.} Representation of Air-Shower and Current status of the Pierre Auger Observatory.}
\end{center}

\section{Objective}
The goal of this work is to observe the effects of direct light in the tanks which conform the SD, as well as their evolution as a function of time. The change in time could provide information on the physical state of each tank. Such information can be also used for studying possible systematic effects in the measurements made by the SD.

\section{The Surface Detector}
Each SD is a cylindrical water tank with $12,000 \ L$ of pure water. It has a cross section of $10 \ m^{2}$, a diameter of $3.6\ m$ and a depth of $1.2 \ m$ (see Figures 2 and 3). The $1660$ tanks of the SD are separated 1.5 km from each other. When an ultra-relativistic CR crosses the tank, it produces Cherenkov light, that is measured by three photomultipliers (PMTs) located on the top of each tank.  Each PMT has $9-in$ of diameter, these are Photonis XP1805. The PMTs are looking to the water, separated a distance of $1.2 m$ from the center of the tank, having a  $120^{\circ}$ of difference in azimuthal angle from each other (see Figure 4).
\begin{center}
\parbox{4cm}{\hspace{-.46cm}
\includegraphics[scale=.1]{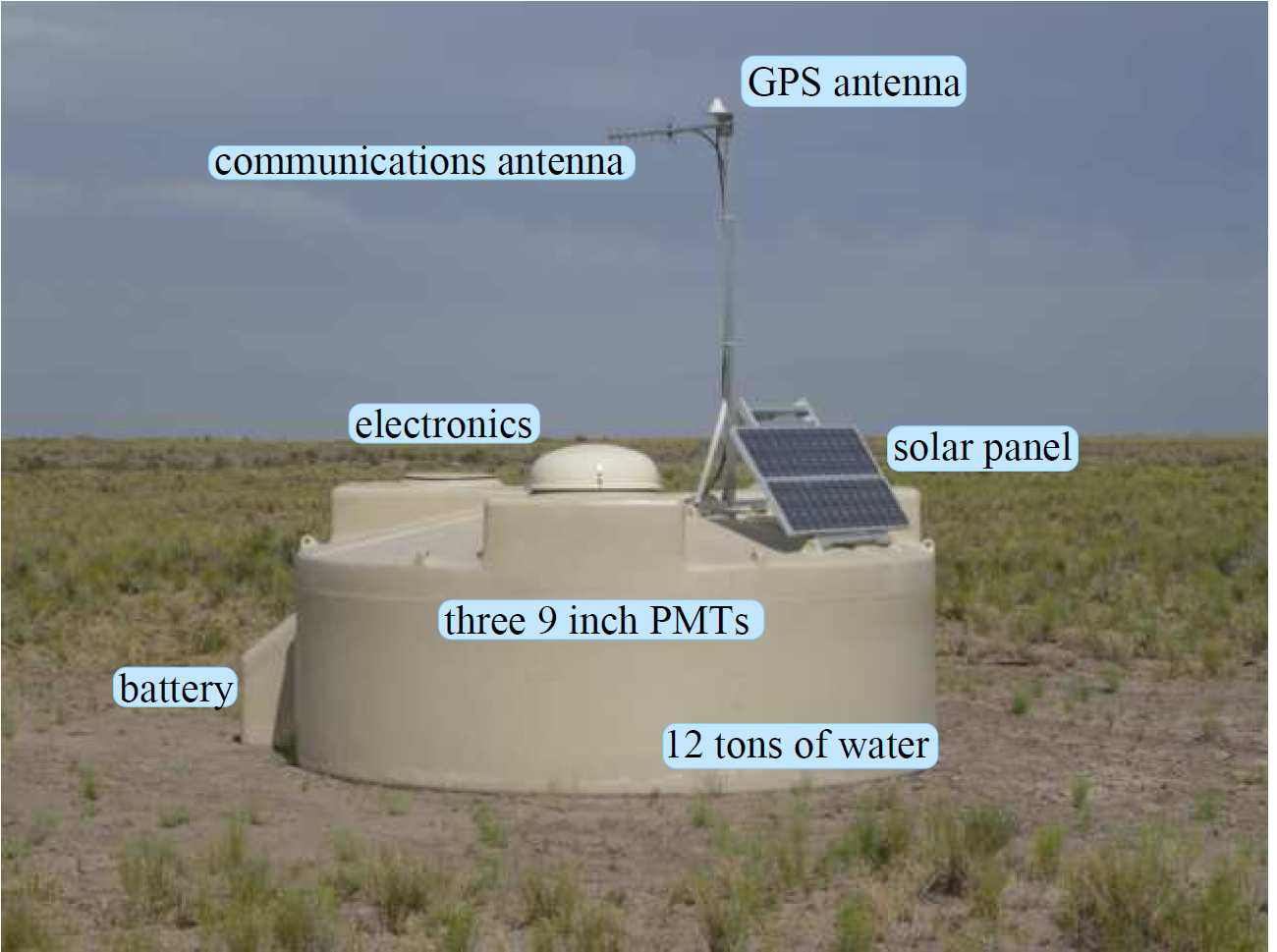}\\
\hspace{-1cm}
{\small\textbf{Figure 2.}\small \ One Surface Detector station.}}
\hspace{.5cm}
\parbox{4cm}{\hspace{0cm}
\includegraphics[scale=.50]{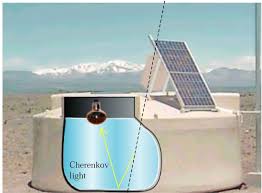}\\
{\small\textbf{Figure 3.}\small \ Particle crossing the tank.}}
\end{center}

\section{Description of the effect of direct light}
The direct light coming to each PMT is conditioned by the azimuthal angle of the arrival direction the particle which produced the light has. If the light arrives with the same azimuthal angle as the one a certain PMT has, then, the measurement made by that PMT will be higher than the one made by the other two. Such effect can be observed in a graph showing the signal as a function of azimuthal angle, which present a different maximum and phase for each PMT, as is shown in Graphs 1 to 4 [2].
\begin{center}
\centering
\includegraphics[scale=.15]{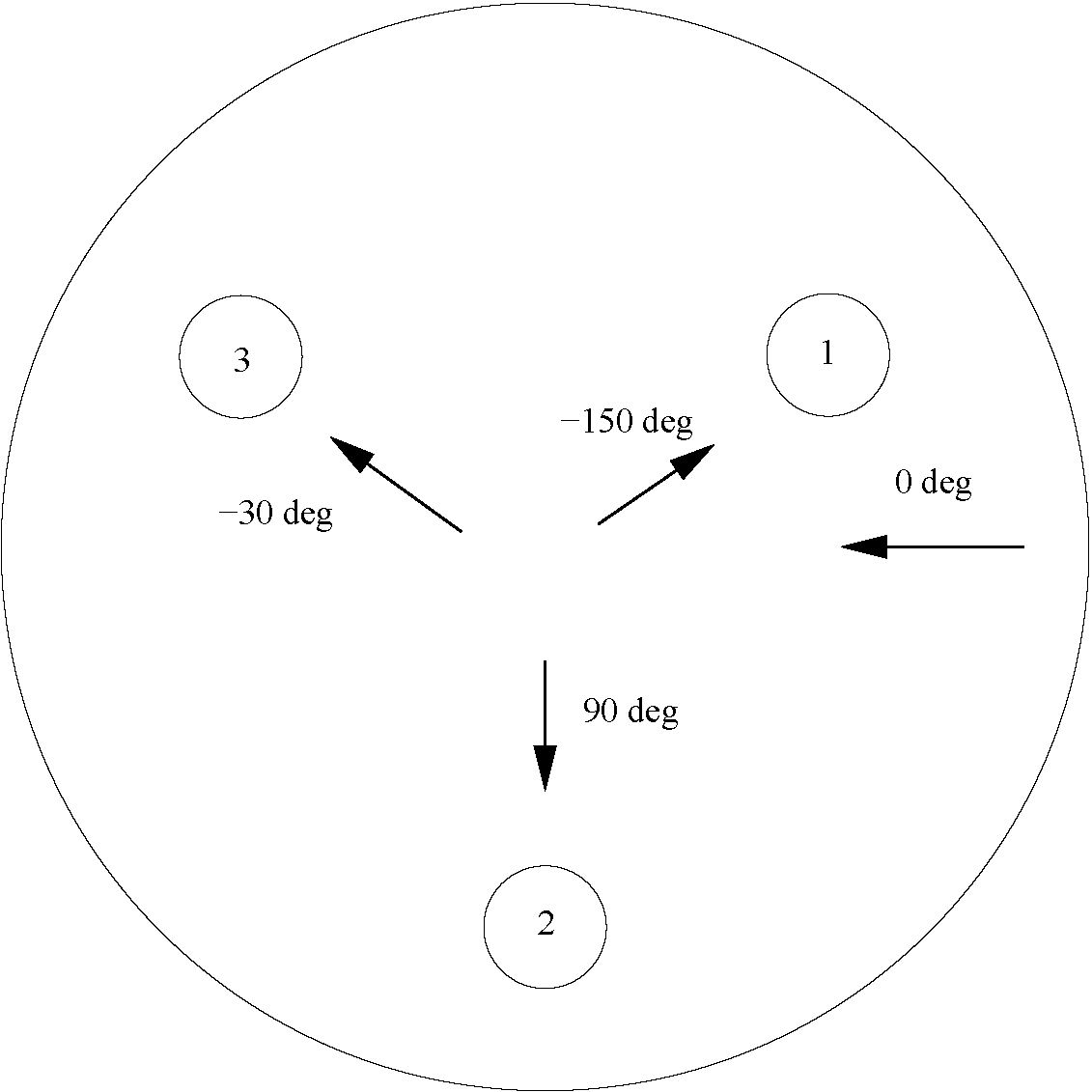}\\
{\small\textbf{Figure 4.}\small \ PMT positions on the top of the tank.}
\end{center}

\section{Description of the Analysis}
A selection of tanks which have been working properly and more or less  continuously over time, during the period from 2004 to 2013, has been made.  Also quality cuts for events detected are requested to guarantee that the measurement corresponds to a well reconstructed event. The results shown in this document are obtained from the analysis of a tank named "Mariana" with number ID 338, which presented an event amount of 5673.\\
Each PMT data is plotted as a function of azimuth angle, and an approximation of the signal from each PMT is performed by adjusting a sine function of the azimuth angle with the form:
\begin{equation}
A\sin(\frac{\pi(x-k)}{180}) + b)
\end{equation}
With $k = 30$ for the PMT1. $k=0$ for PMT2 and $k=30$ for PMT3.

\begin{center}
\centering
\includegraphics[scale=.7]{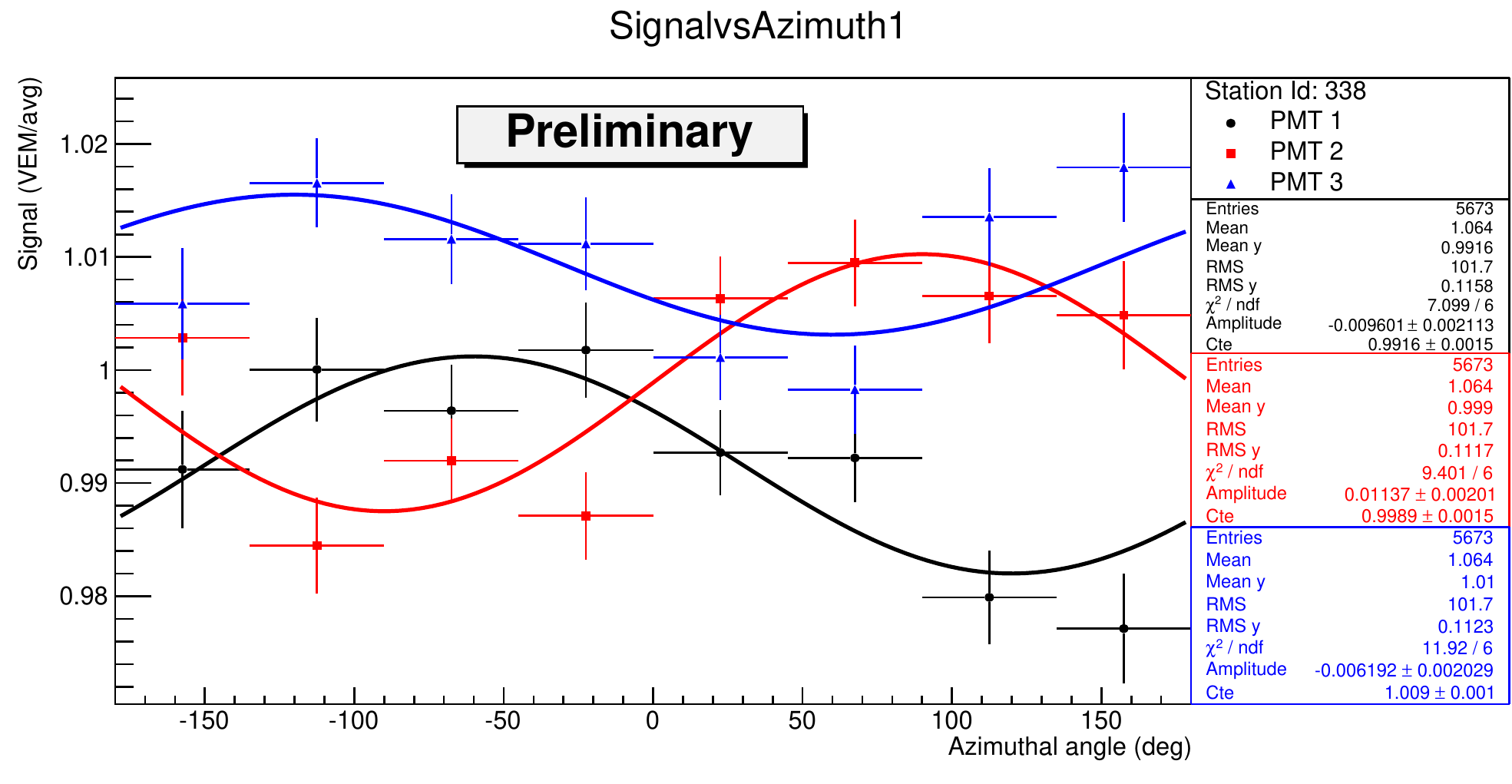}\\
{\small\textbf{Graph 1.} \ Amplitude 0$^{\circ}$-60$^{\circ}$.}
\end{center}

\noindent Three zenith angle ranges, having the same solid angle, are considered: $0^{\circ}-34^{\circ}, 34^{\circ}-48^{\circ}\ \text{and} \  48^{\circ}-60^{\circ}$. The corresponding results, showing the azimuthal effect are shown in Graphs 2 to 4.
\begin{center}
\centering
\includegraphics[scale=.38]{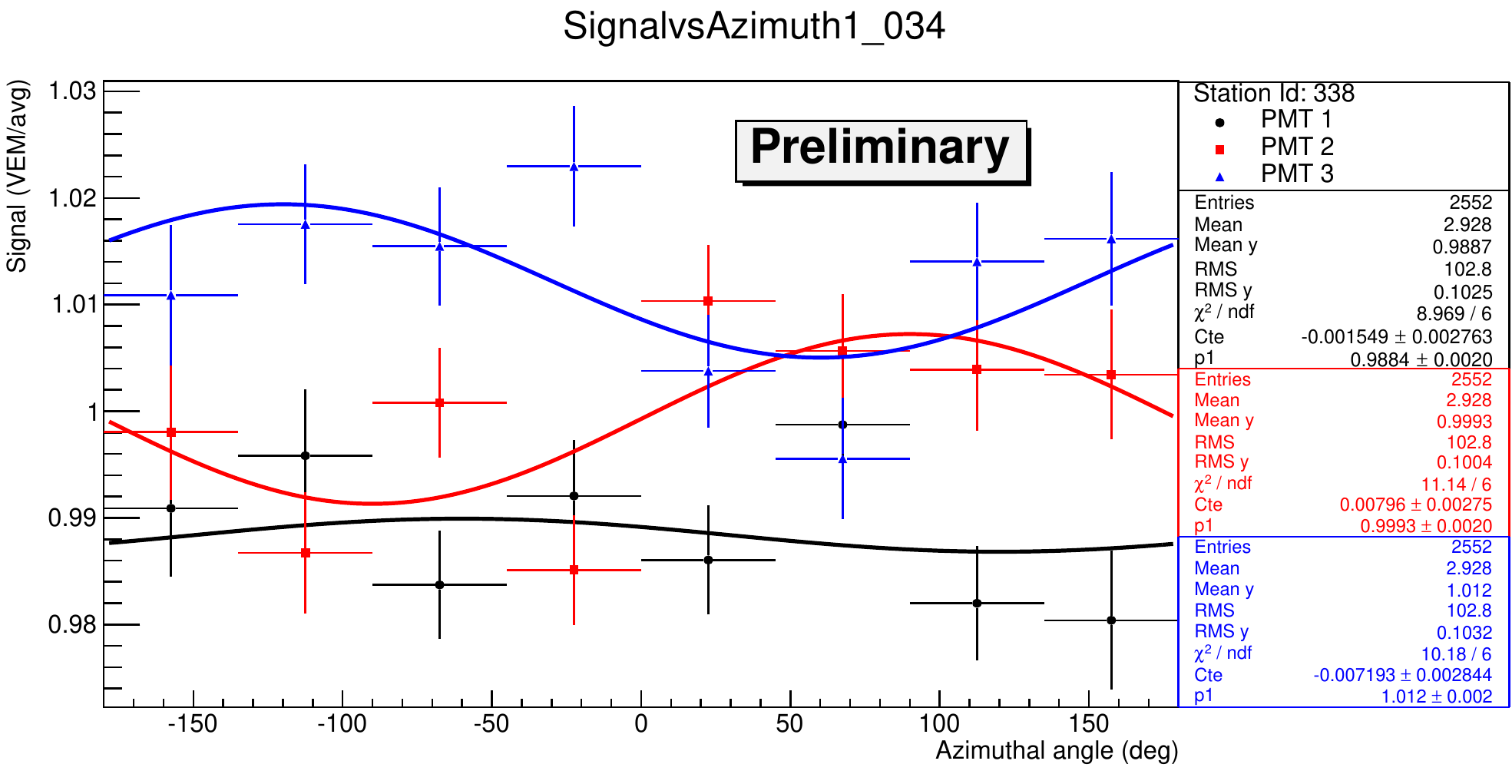} \hspace*{.5cm} \includegraphics[scale=.38]{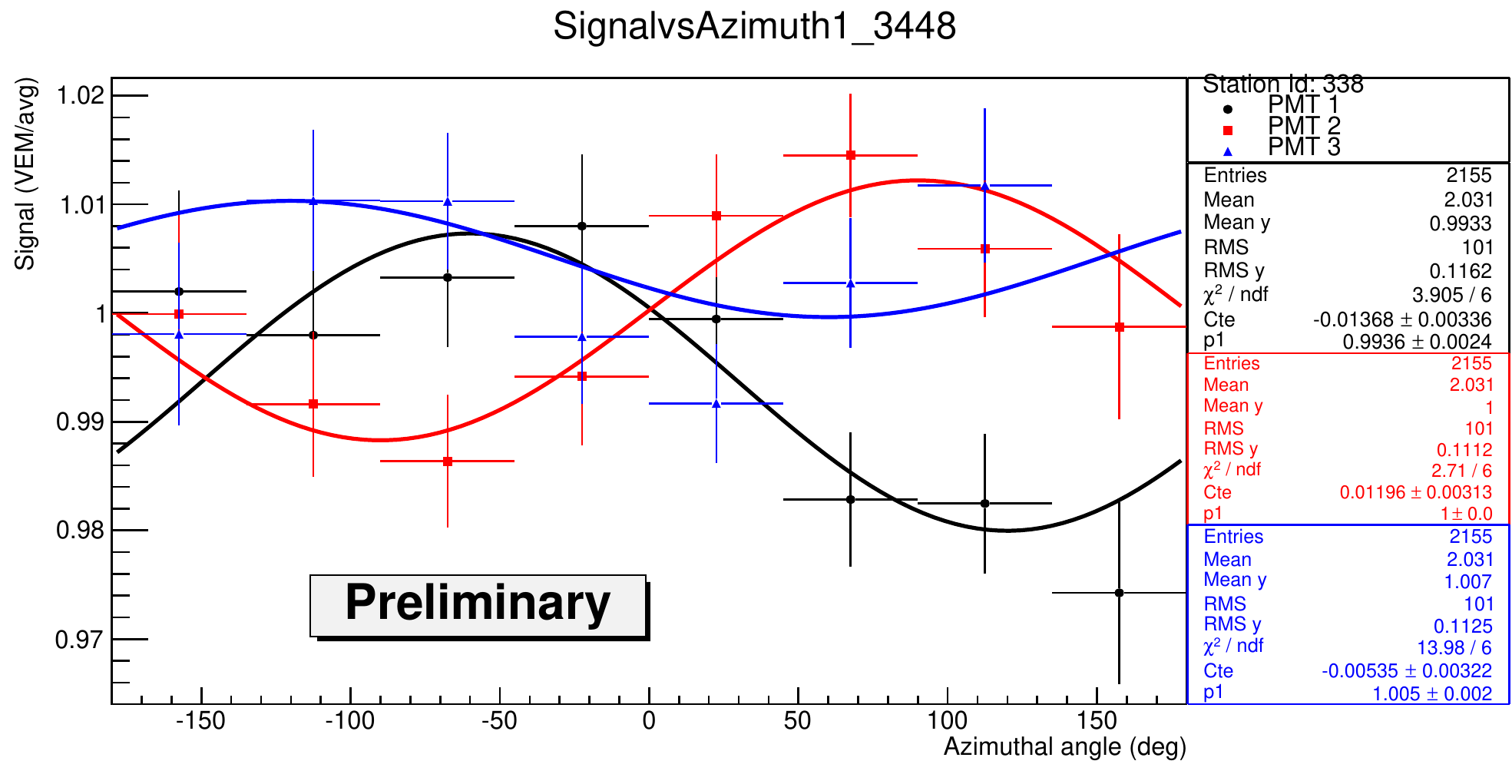}\\
{\small\textbf{Graph 2.} \ Amplitude 0$^{\circ}$-34$^{\circ}$.}\hspace*{4cm} {\small\textbf{Graph 3.} \ Amplitude 34$^{\circ}$-48$^{\circ}$.}\\
\includegraphics[scale=.38]{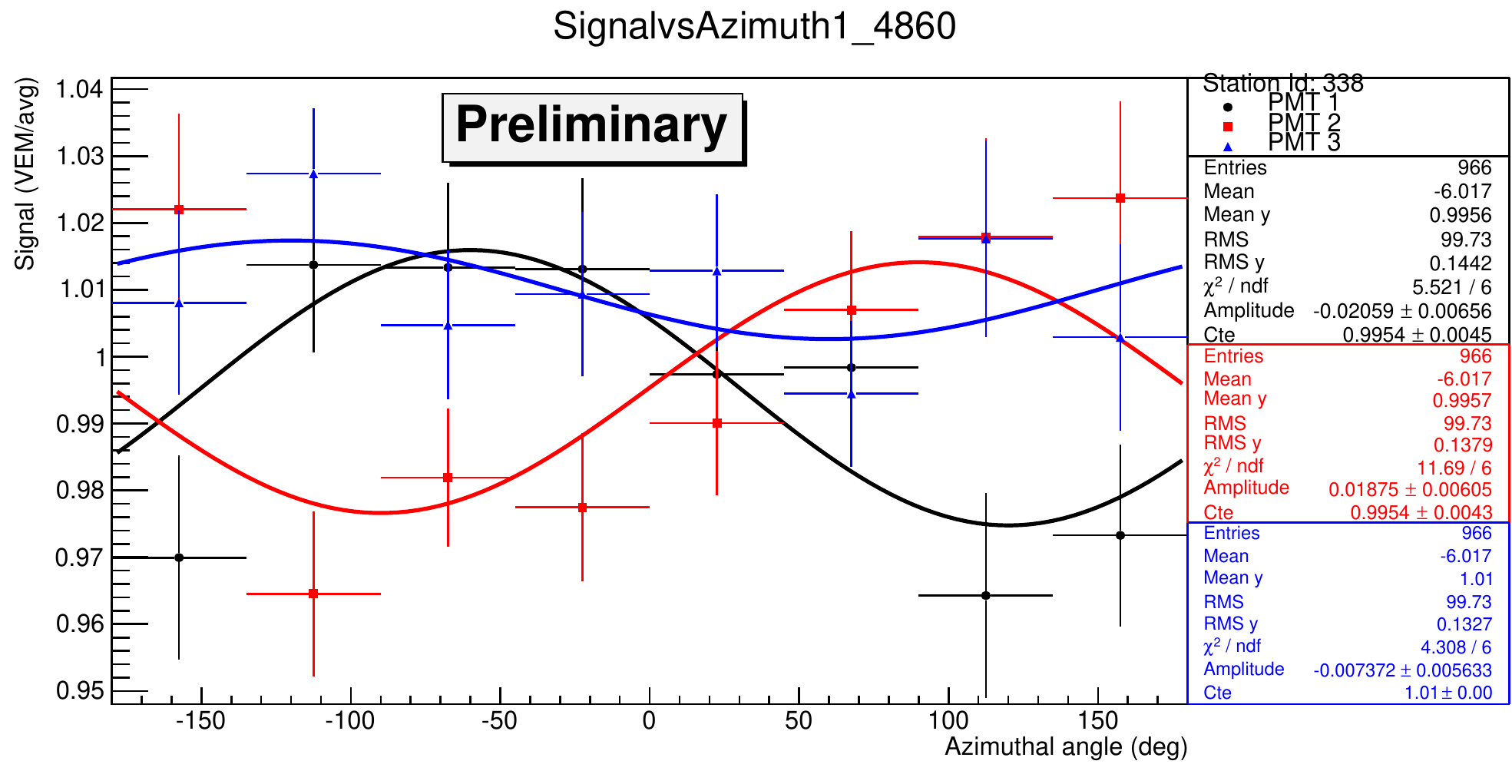}\\
{\small\textbf{Graph 4.} \ Amplitude 48$^{\circ}$-60$^{\circ}$.}
\end{center}
The amplitudes of each PMT are plotted as a function of zenith angle. The corresponding line is fitted with a function of the form:
\begin{equation}
y = mx + b
\end{equation}
Results for PMT 1, PMT 2 and PMT 3 are shown in Graphs 5 -7.
\begin{center}
\hspace*{-1cm}\includegraphics[scale=.45]{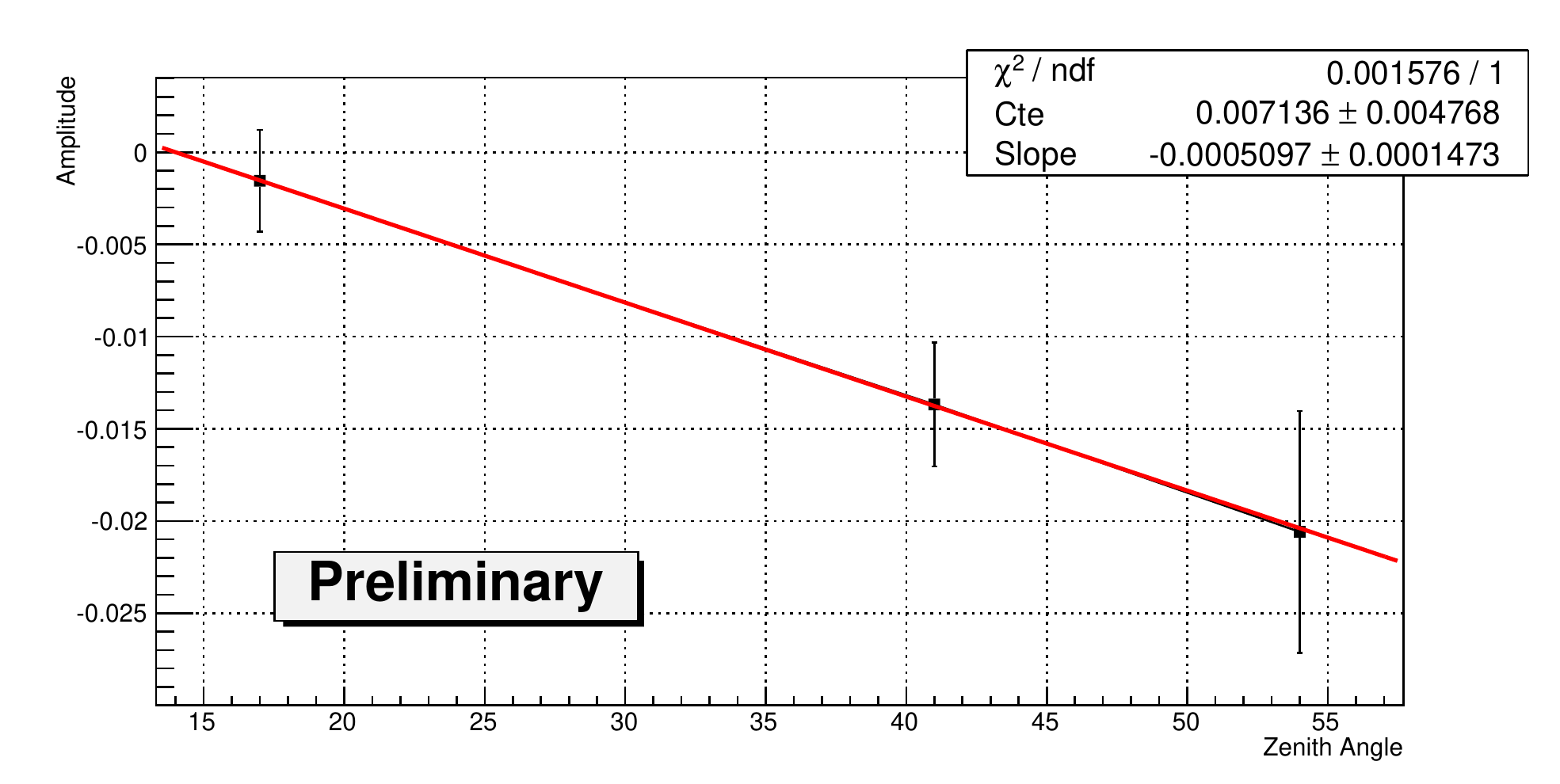}\includegraphics[scale=.45]{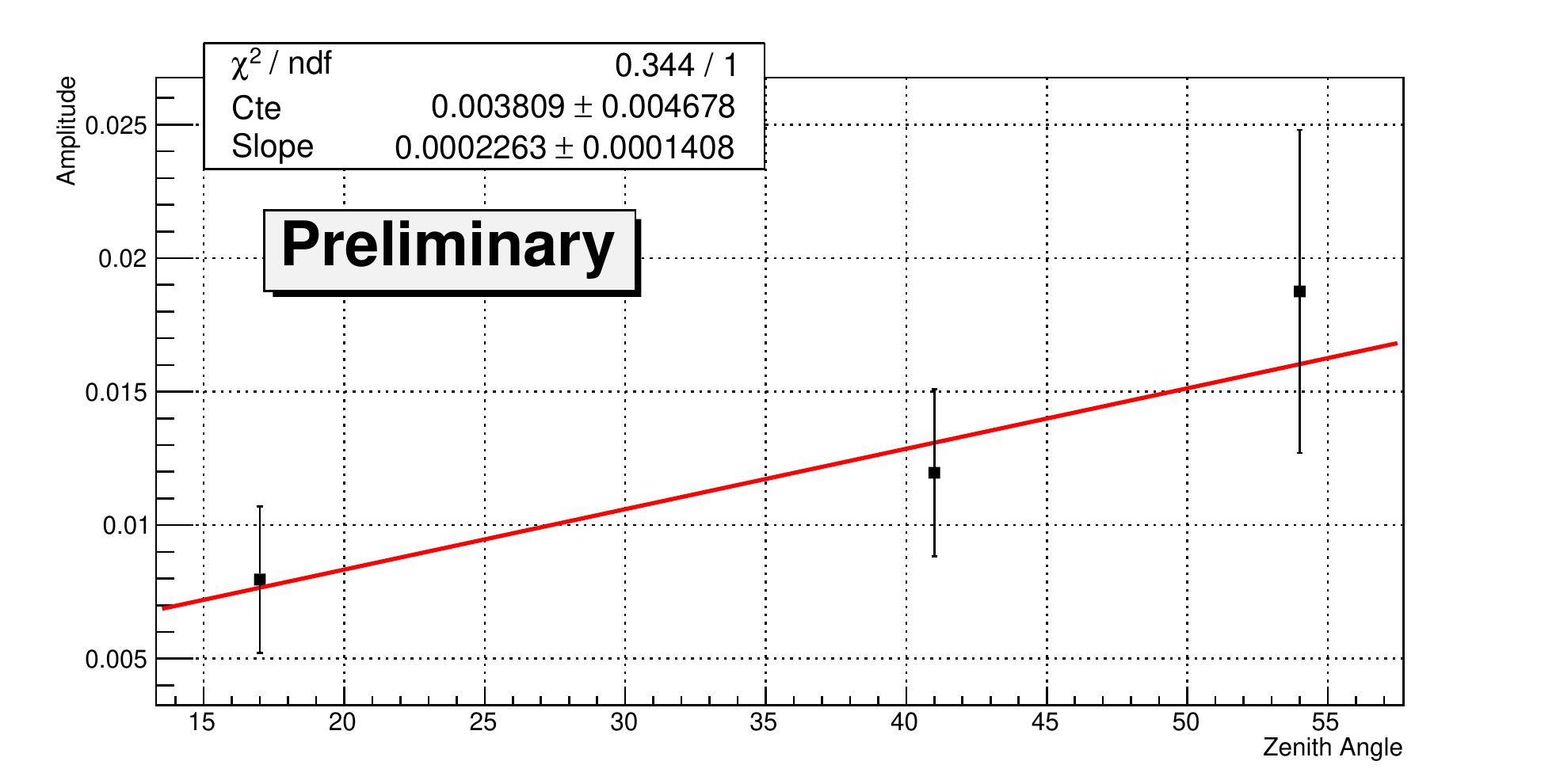}
{\small\textbf{Graph 5.}\ Amplitude adjustment for the PMT1.}\hspace*{1cm}{\small{\textbf{Graph 6.}\ Amplitude adjustment for the PMT2.}}\\
\includegraphics[scale=.45]{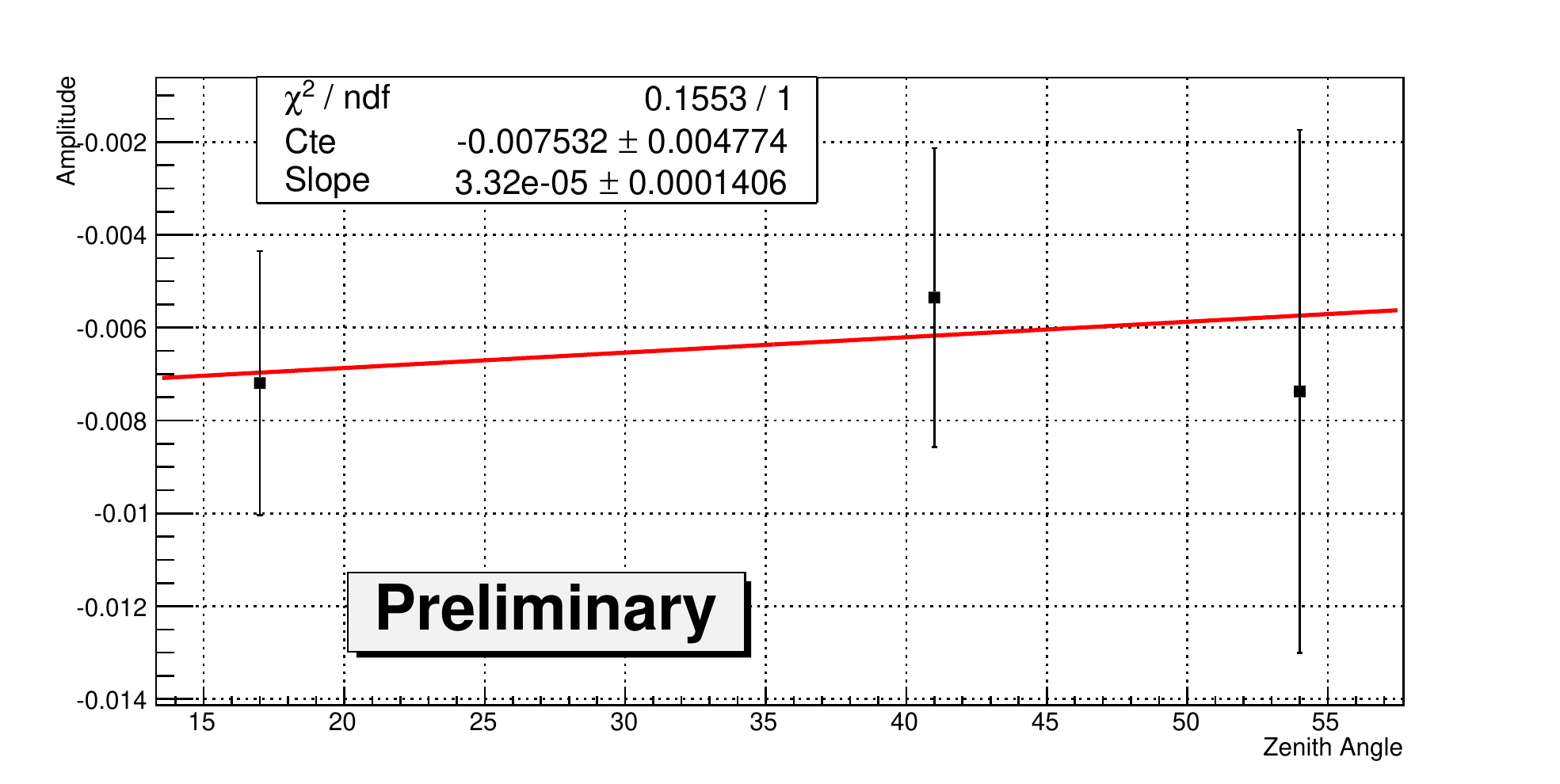}\\
{\small\textbf{Graph 7.}\ Amplitude adjustement for the PMT3.}
\end{center}
In order to obtain only one characteristic line for each detector, the amplitudes for each PMT at a certain zenith angle are averaged and plotted as shown in Graph 8. Then a new linear fit is performed resulting in the corresponding characteristic line.

\begin{center}
\includegraphics[scale=.6]{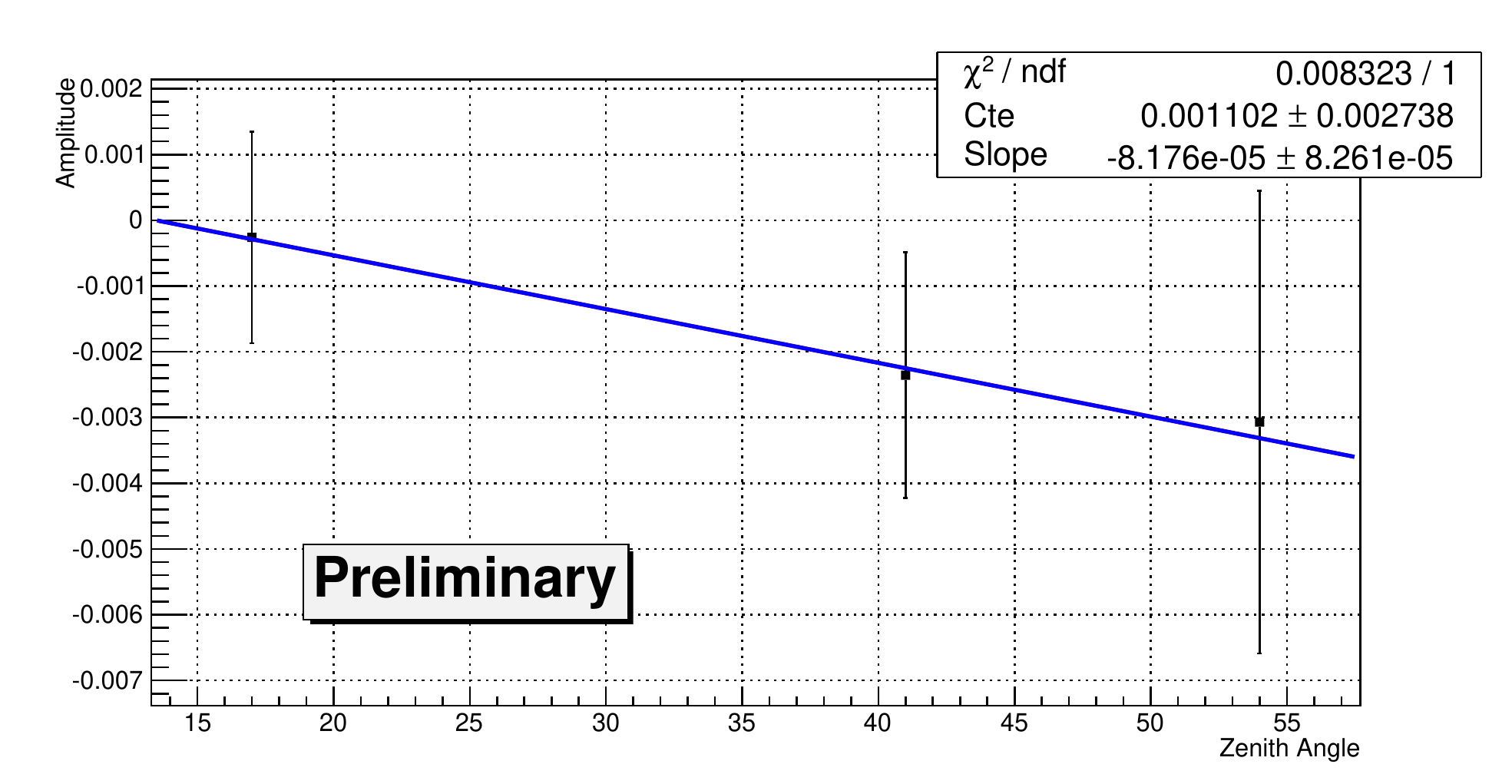}\\
{\small\textbf{Graph 8. \ Characteristic line of the station.}}
\end{center}
Once the characteristic line of the station was obtained for the whole period of time, its change in time is to be explored. The same study has to be done for three different periods in order to compare the corresponding characteristic lines. This is work in progress.
\section{Conclusion}
From graphs 1-4, it can be seen that the effect of direct light is effectively present for each PMT of of one station in the SD. A characteristic line, for the station studied was found. The next step is to explore signals of the same tank for different time periods in order to study a possible change of the characteristic line as a function of time. If some change is found, its impact in the measured signal has to be studied. The same kind of study must also be performed for other tanks.
\section{Aknowledgements}
\begin{itemize}
\item To CONACyT for the support given by the project CB 243290.
\item To L'ORÉAL Mexico, CONACyT, UNESCO, CONALMEX and AMC fort the support granted through the Fellowship for Women in Science 2014 to Karen Salom\'e Caballero Mora.
\end{itemize}

\section{References}
\begin{itemize}
\item[1]\textsc{The Pierre Auger Collaboration, arXiv:1604.03637v1, astroph-IM, The Pierre Auger Observatory Upgrade Preliminary Design Report (2016).}
\item[2] \textsc{K.S. Caballero-Mora, L. Nellen and J.F. Valdes-Galicia.}\textit{ Azimuthal signal variations in the engineering array of the Pierre Auger Observatory}. Revista mexicana de Física Ag 2008 54 (4) 306-313

\end{itemize}

\end{document}